\documentclass{article}
\usepackage{url}
\usepackage{enumitem}
\usepackage[a4paper, total={5.5in, 8in}]{geometry}
\usepackage{hyperref}

\usepackage{amsfonts}
\usepackage{amsmath,amssymb}

\usepackage[numbers,sort&compress]{natbib}
\usepackage{authblk}
\hypersetup{
    colorlinks,
    linkcolor={red!50!black},
    citecolor={blue!50!black},
    urlcolor={blue!80!black}
}

\usepackage{todonotes}
\newcounter{todocounter}

\usepackage{xspace}
\usepackage{float}
\usepackage{boxedminipage}

\newif\ifshowcomment
\showcommenttrue
\ifshowcomment
\newcommand{\todos}[1]{\stepcounter{todocounter}\textcolor{red}{{ [[(\thetodocounter) TODO: #1]]}}}
\newcommand{\SW}[1]{{\footnotesize\color{olive}{[Sarisht: #1]}}}
\else
\newcommand{\todos}[1]{}
\newcommand{\SW}[1]{}
\fi

\newcommand{\protocol}{\ensuremath{\Pi}\xspace}

\newtheorem{definition}{Definition}
\begin{document}

\title{Differentially Private aggregate hints in MEV-Share}

% \author{Jonathan Passerat-Palmbach, Sarisht Wadhwa\\
% \email{jonathan@flashbots.net, sarisht.wadhwa@duke.edu}\\\affiliation{%
%   \institution{Flashbots Research}
%   \country{UK}}}
\author[1]{Jonathan Passerat-Palmbach\thanks{jonathan@flashbots.net}}
\author[2]{Sarisht Wadhwa\thanks{sarisht.wadhwa@duke.edu}}
\affil[1]{Flashbots}
\affil[2]{Duke University}

\date{}
\maketitle
\begin{abstract}
    Flashbots recently released MEV-Share to empower users with control over the amount of information they share with searchers for extracting Maximal Extractable Value (MEV). Searchers require more information to maintain on-chain exchange efficiency and profitability, while users aim to prevent frontrunning by withholding information. After analyzing two searching strategies in MEV-Share to reason about searching techniques, this paper introduces Differentially-Private (DP) aggregate hints as a new type of hints to disclose information quantitatively. DP aggregate hints enable users to formally quantify their privacy loss to searchers, and thus better estimate the level of rebates to ask in return. The paper discusses the current properties and privacy loss in MEV-Share and lays out how DP aggregate hints could enhance the system for both users and searchers. We leverage Differential Privacy in the Trusted Curator Model to design our aggregate hints. Additionally, we explain how random sampling can defend against sybil attacks and amplify overall user privacy while providing valuable hints to searchers for improved backrunning extraction and frontrunning prevention.

\end{abstract}
\thispagestyle{plain}
\pagestyle{plain}
\section{Introduction}

Maximal Extractable Value (MEV) has proven to be a lucrative domain, generating over 200k Eth \cite{flashbotsdashboard} since September 2022 by exploiting information released by traders while interacting with decentralized exchanges \cite{ DBLP:journals/corr/abs-1904-05234}. Some MEV extraction strategies like arbitrage and liquidation are crucial in market efficiency, as they help balance prices across multiple markets. Conversely, strategies such as frontrunning and sandwiching are deemed detrimental to the decentralized exchange ecosystem, as they exploit traders' transactions, resulting in worse execution prices \cite{ DBLP:journals/corr/abs-1904-05234}. This creates a dilemma for traders: revealing information improves market efficiency but may lead to worse trades. Users who employ these strategies, both beneficial and harmful, are referred to as searchers, and their strategies are collectively termed searching.

Addressing this dilemma has been discussed in the Ethereum community and beyond. Various ideas have been proposed to solve the problem. While some approaches initially proposed to prevent MEV extraction completely, recent proposals converge towards enabling some form of controlled extraction, commonly called programmable privacy. In an all-or-nothing approach, we have seen multiple proposals enabling searchers to backrun transactions while preventing frontrunning \cite{mpcbackrun, mevblocker} in a fully blind way. However, these techniques have limitations in trust assumptions and are impractical since the searcher's strategy would need to be publicly posted for verification, revealing crucial information. 

% When we look at the difference between the two types of information usage (both good and bad); we see that we wish to enable transactions that can be placed after a trade, but not before the trade. This gives us another lever to play with since the transactions on a chain like Ethereum can be arbitrarily programmed to read the current state of the blockchain, like getting the current execution price of a certain asset before programming the trade. Although this works in theory, the on-chain cost to the searcher is very high, making it infeasible to extract the information completely from the chain. We will talk about this again in section \ref{sec:search:simplearbitrage}.

With a more granular approach in the programmable privacy direction, MEV-Share\cite{mevshare} was launched to enable users to disclose some fields of their transactions as hints. Searchers can exploit these hints to extract MEV more efficiently and offer rebates to users for the information they share. A central actor, the matchmaker, connects users and searchers, ensuring no front-running bundles are submitted.

This new way of searching relies explicitly on private channels/relays and on the matchmaker not to reveal all the information to the searchers, thus creating a search space for the searchers to "guess" how to extract profits from the trade. MEV-share's take on programmable privacy requires the users to estimate whether the information they are leaking via hints is worth the rebate amount searchers will pay them. Users need to quantify their loss of privacy while having no guarantees on how much information they are revealing to elaborate searchers.

In this paper, we introduce a new type of hints, the \textbf{Differentially-Private (DP) aggregate hints}, that enable users to adopt a new type of fully quantified information disclosure. Using DP aggregate hints, users can quantify their privacy loss and thus better estimate the rebate they should ask in return. At the same time, searchers have access to new information from the matchmaker that can help improve their strategies and the amount of MEV they can extract from the system.

We describe the properties and privacy loss incurred in the current form of MEV-share in Section \ref{sec:searching_mev_share}. Our new proposal to enhance MEV-share with DP aggregate hints is presented in Section \ref{sec:aggregate_hints}. We conclude this study by discussing the technical details regarding DP that must be considered to deploy our approach in practice (Section \ref{sec:discussion}).

% we try to improve the search space for searchers in the last mentioned method. In MEV-Share, the users define the search space for the searchers to search for profits. The users decide how much information they wish to reveal to the searchers. In addition to the search space reduction by revealing certain information, we also add additional aggregated hints. Any information completely released is a part of the aggregate of that information, however, a user can choose to not reveal some information while being a part of the aggregate set.

% In conclusion, the evolving landscape of decentralized finance and the search for MEV strategies necessitates balancing market efficiency with privacy concerns. Various techniques have been proposed, and MEV-Share provides a promising approach with aggregated hints and differential privacy to empower searchers while safeguarding user information.

\section{Background}
\paragraph{MEV\cite{DBLP:journals/corr/abs-1904-05234}} Decentralized exchanges on different consensus systems create inefficiencies as their numbers increase. MEV, or Maximal Extractable Value, represents the profit extracted from exchanges by the underlying consensus, either within the same block or distributed across multiple blocks on the same or different chains. As mentioned in Introduction, we refer to those seeking MEV as "searchers."

\paragraph{Flashbots Protect\cite{Fbprotect}} 
\label{sec:background:fbprotect}
Flashbots introduced Protect as an RPC endpoint to prevent users' transactions from being frontrun by MEV bots in the public mempool. It guarantees that only some searchers trusted by the relay (based on reputation) can look at a transaction and then prevent those bundles that frontrun the transaction from reaching the block-builders. This also adds another functionality: a bundle submitted by the searcher is invalid unless the complete bundle goes on-chain. For the bundle to be accepted on-chain, the block builder auctions the block space to the highest-paying searchers. Consequently, a searcher is allowed to submit multiple bundles for the same trade, and is ensured that only one makes it on-chain. However, since this is a reputation-based model, the searcher is disincentivized to spam the system to create the best bundle. To represent this in practice, we assume that the number of bundles a searcher can submit is rate-limited in the present version of Flashbots' infrastructure.

\paragraph{MEV-Share\cite{mevshare}}
 MEV-Share was recently introduced as a part of Flashbots Protect to hide user transactions from searchers. The main problem with using a relay was that the searcher could submit an individual frontrunning transaction directly to the block builder, paying more value than the bundle itself. This would allow the frontrunning transaction to be placed before the bundle, allowing a sandwich without being sent through the protect RPC. Thus, MEV-Share aims to enable users to program their transactions' privacy to disallow the searcher from being able to form an unbundled sandwich. They introduce a trusted party called matchmaker. The matchmaker exposes to searchers the data fields the users opted to disclose as part of the programmable privacy feature. This piece of information is called a hint. Searchers try to maximize the extracted MEV using hints and specify a rebate (in \%) that goes back to the user. The matchmaker collects the bundles created by the searchers and adds private transactions based on data provided by the searchers (such as a hash of the transaction). The bundle with the maximum kickback to the user is forwarded to the block builder by the matchmaker.

\paragraph{Differential Privacy} Differential Privacy is a concept introduced by \cite{dwork2006calibrating} to balance data utility and privacy. It quantifies the privacy guarantees a mechanism can provide. It says that if there are two neighboring datasets $X$ and $X'$ where neighboring means that they differ in exactly one element (add, delete, or replace), then the output from the mechanism follows a distribution with the following property: 

\begin{definition} {($(\epsilon, \delta)$-DP)}
\label{eq:dpdef}
A randomized mechanism $M$ satisfies $(\epsilon, \delta)$-differential privacy if for all pairs of neighbouring datasets $X$ and $X'$ and all sets of outputs $S$ we have:
$$Pr[M(X) \in S] \leq exp(\epsilon)\cdot Pr[M(X') \in S] + \delta$$
\end{definition}

For future analysis, we would represent neighboring datasets $X$ and $X'$ as $X \sim X'$

\paragraph{Noise Characteristics} Differential Privacy is achieved by adding noise to the query. However, not all types of noise can satisfy differential privacy. Two major noise distributions and their requirements to maintain differential privacy are discussed here: 
\begin{itemize}[leftmargin=*]
    \item Laplacian Noise: Noise that follows the Laplacian distribution, with mean $(\mu) = 0$ and the scaling factor $(b) = \frac{\Delta_1}{\varepsilon}$
    \begin{equation}
    f(x|\mu,b) = \frac{1}{2b} \exp\left(-\frac{|x-\mu|}{b}\right)
\end{equation}
    where $\Delta_1$ is the $l1-$sensitivity of the mechanism/query being run. 
    The resultant mechanism is known to be $(\varepsilon, 0)-$Differentially Private
    \item Gaussian Noise: Noise that follows the Gaussian distribution, with mean $(\mu) = 0$ and the variance $(\sigma)=\sqrt{2\log(1.25/\delta)}\frac{\Delta_2}{\varepsilon}$
    Gaussian noise allows adding less noise and enables more advanced composition between mechanisms. The guarantees it offers are, however slightly weaker. For clarity, we will define all mechanisms as pure Laplacian DP in the rest of the paper, but our proposal could also support approximate Gaussian mechanisms. 

\end{itemize}

% Any number of mechanisms can be composed. i.e., if a searcher receives k pieces of information from various queries made, all of which are $(\varepsilon,\delta)$-DP, then the result is also differentially private, with parameters $(\epsilon' = \sqrt{2k \log(1/\hat{\delta})} \epsilon + k  \epsilon  (\exp{(\epsilon)} - 1), \delta' = k\delta+\hat{\delta})$.

The main property of DP is that it is always safe to post-process results from a DP mechanism. Hence, searchers can apply arbitrary computations to the released results without being able to reverse the privacy protection. 

The amount of noise added to a mechanism is determined based on the sensitivity of the function to protect. The sensitivity of a function represents the maximum difference ($l1$ or $l2$ depending on the type of noise added) between the function evaluated at neighboring points in a dataset. 
% \section{Related Work}
% -DP papers

\section{Searching in MEV-Share}\label{sec:searching_mev_share}

The role of searchers is to find as much MEV as possible. However, in MEV-Share, the searcher does not have all the information required to create a bundle accurately and is only provided hints on how to build a bundle. When the bundle is created, the matchmaker must insert the complete transaction and send it to the block builder. As a searcher, without the knowledge of the content of the transaction, there exist two ways of gaining information about the transaction: either program that inside a contract, information is extracted from relevant on-chain states of contracts like the Uniswap v2 contract; or brute force search the entire space of contracts with various bundles. While the former gives us complete information to backrun, it takes much more gas costs than the brute force search. This section provides more details on the two ways a searcher can utilize to gain information and describes the privacy loss incurred in the current MEV-share protocol.

To describe searchers' strategies more formally, we define a trade as a tuple $(P=(T_b, T_s), \protocol, val)$, where $T_b$ is the token being bought, $T_s$ is the token being offered in return, \protocol is the address of contact being accessed and $val$ represents the amount being traded. $P=(T_b, T_s)$ is also jointly referred to as the pair being traded. Now, a user sending a transaction could potentially try to trade multiple assets in a single transaction, and thus we define the transaction sent by the user to be a list of all the trades. 

Finally, we assume the existence of a matchmaker party that is trusted for the privacy of the transaction and the integrity of the information it provides. We further assume an ideal block builder is trusted to conduct a fair auction of the block space to all the searchers, i.e., not steal the searcher's bundles or cooperate with any searcher.

\subsection{A backrunning contract}
\label{sec:search:simplearbitrage}

In this section, we utilize the hints provided by the user and write a program to derive the rest on-chain before the execution of the transaction. Assume that the user releases the Protocol/Pair combination. We use the simple blind arbitrage bot \href{https://github.com/flashbots/simple-blind-arbitrage}{simple blind arbitrage bot}\cite{Flashbots} as the inspiration for this contract. Similarly, if the user releases only the pair but not the protocol, then the on-chain contract checks whether the state of the pair has changed in various protocols and, if it has, appropriately finds the change in its state. 

As a detailed example, let us assume the user releases both the Protocol and the Pair $( \Pi, P)$. For the pair, an offline price call requests a price from other protocols and finds the protocol with minimum price ($\Pi_M$) ($\Pi \neq \Pi_M$). These three and the current liquidity of $\Pi$ are input into the contract through a function. This function is what is being called in the backrun transaction. The function calls $\Pi$ for its current value so that the contract can infer the amount traded inside the user’s transaction. A greedy backrun is performed with an asset exchange between $\Pi$ and $\Pi_M$.

The contract (pseudo-code available in appendix - Fig. \ref{fig:contractbackrunning}) would serve as the baseline for comparison when no aggregate information is provided, i.e., a condition where only the protocol and pair have been released, and the direction and amount are withheld. It is imperative to note that this still can be frontrun and sandwiched using statistical frontrunning. However, such attacks are inevitable and are beyond the scope of this study.

\subsection{Brute Force Search Space}
\label{sec:search:bruteforce}

As mentioned in the background, the searcher can send various bundles for the same transaction. Even though most of them are not profitable or even valid, the block builder would choose the most profitable bundle to get the most profit for itself (More extracted value implies more profit for the block builder since it gets a \% cut from the extracted value). This means that the searcher could create a lot of bundles and send them all to the matchmaker to forward to the block builder. Choosing the best MEV extracting bundles is thus forwarded to the block builder. Recall that we assumed the bundles received from a searcher and the bundles the matchmaker sends to the block builder are rate-limited (Sec \ref{sec:background:fbprotect}, (Flashbots Protect)) (i.e., a block builder would only receive less than a threshold amount of bundles from the matchmaker). 

Further, here the search space can be searched heuristically by making statistical decisions, i.e., which transactions are more likely to appear due to external (off-contract) reasons. Let us assume that the statistical analysis of the chain returns some probability distribution of the values: the protocol, pair, and amount. 

Comparison for both these methods is presented in appendix \ref{comparison}

\subsection{Privacy loss in MEV-Share}
According to the specification released for MEV-Share\cite{mevshare}, programmable privacy is offered to users, i.e., users can choose the limit of information released to searchers. The more information available to the searcher, the more MEV it can extract, and thus the higher it can bid to return to the user. However, extracting MEV that impacts the user trade rate is easier if all the information is available, like sandwich attacks \cite{hft}. 

We first ask the question: what information is enough for a searcher to accurately determine a trade and profit from the trade by sandwiching the transaction that contains the trade? We narrow down three pieces of information to be essential for the searcher: the protocol being interacted with (Uniswap v2\cite{adams2020uniswap}, curve\cite{curve}, Uniswap v3\cite{adams2021uniswap}, etc.), the pair of tokens (USDC, Eth, WBTC, etc.), and the amount being traded constitute enough information for a searcher to reconstruct the entire transaction. As stated in the background, the aim is to give searchers enough information to backrun the transaction to make the auction profitable while preventing a negative user experience.

Even when using Flashbots Protect, users are still vulnerable to frontrunning-based attacks such as sandwiches. This is because when all the necessary information is available to a searcher, it can split the sandwich into two parts: the frontrun transaction and the backrun bundle. Here, the frontrun transaction is all the trades that the searcher would have done in a sandwich attack before the actual victim trade, and the backrun bundle would contain the victim transaction followed by the trades after the victim transaction. The frontrun transaction pays a high transaction fee to be placed before the backrun bundle (preferably right before the bundle). This way, the searcher could extract sandwich profits without submitting invalid inputs to the Flashbots Protect RPC. 

When we switch to MEV-Share, this problem is somewhat addressed since the complete transaction to frontrun is hidden. This means that the searcher would need to search for the frontrunning transaction. And since the searcher no longer has the advantage of Flashbots Protect, the submitted frontrunning trade can make it on-chain without the bundle being accepted. Since both trades are in the same direction on the same protocol, this would mean that another searcher with an optimal backrunning would extract more MEV than in the case where the unbundled sandwich was not attempted. Thus, the user needs to be careful in what information it discloses to the searcher via MEV-Share and how it can reduce the search space only for backrunning transactions.

\section{Aggregate hints}\label{sec:aggregate_hints}

We introduce the concept of aggregate hints, in which when the user does not want to release some aspect of a trade, it can opt in for aggregate hints, which releases information about all the transactions opting in for such aggregate hints. Now, the question arises how are the aggregate hints useful? To answer that, we first examine what kind of aggregates are possible. Although we can't exhaustively list all the available aggregate functions, an example would be to aggregate the amount traded. For all the trades where the user opts in to give aggregated statistics to a particular protocol, the sum of inputs could be released to the searcher as a hint. This helps the searcher narrow down the search space for all backrunning bundles (as well as frontrunning bundles, though quantification of the same is important as well). Another type of aggregation could be a count of transactions that satisfy a property. Let's say the user releases which pair he wishes to trade but not the protocol that it interacts with. The user can allow the transaction to be a part of a count query when the count of other protocols exceeds a user-determined threshold. This helps the searcher determine how many transactions in the block interact with different protocols to estimate how many of each transaction should be used. Thus, aggregate queries can help reduce the search space for the searcher, thus requiring fewer queries to the matchmaker and the block builder. 

The idea proposed can support various other types of aggregate queries, which would help to reduce the search space for the searchers or provide better than real-world statistics heuristics; however, for the paper, we only present a few of them. We note that these queries can also be combined to obtain new results: a count and a sum on the same input parameters allow the requester to learn the mean at no extra privacy cost for the system.

\subsection{Trusted Curator Model}

MEV-Share has a player called the matchmaker ($M$) that takes users' transactions and releases only certain information for searchers to extract MEV. Since the player is already trusted for the programmed transaction's privacy, we also give it the role of the Trusted Curator for differential privacy. The Trusted Curator aggregates the raw data and applies a global noise to the mechanism computed over this data, according to the sensitivity of the function. We divide the protocol into two phases:

\noindent\textbf{Input:} In this phase, the users ($U_1, \cdots, U_n$) submit their transactions to the matchmaker $M$. The matchmaker parses the relevant information as \\
$(tx_i, data_i, [(aggCond_i^1, agg_i^1),\cdots,(aggCond_i^k, agg_i^k)])$ for each user $U_i$ that submitted the transaction $tx_i$, releases data $data_i$ in the clear to the searcher, and releases aggregate of $agg_i^k$ when the condition $aggCond_i^k$ is satisfied. For example, a user may contribute to the aggregation of the count of UniswapV2 interactions once at least ten transactions opt to do so and at least ten others opt not. Alternatively, a user could reveal the aggregate amount.

\noindent\textbf{Process:} In this phase, the matchmaker $M$ runs the aggregate queries, and broadcasts  
    $(data_i, [(aggCond_i^1, agg_i^1),\cdots,(aggCond_i^k, agg_i^k)$  $ \forall aggCond_i^k = True])$ where $agg^k$ is the appropriate aggregate function applied on all $tx_i$ for which $aggCond_i^k = True$
    In this work, we will show a few aggregate functions: \textit{countOf} and \textit{sumOf}, where countOf would return the number of transactions. As the name of the paper suggests, these queries would have a differentially private response since we do not want to leak enough information such that the frontrunning search space reduces to within a threshold, but reduce the backrunning search space such that the brute force search space for backrunning becomes much more efficient.

\subsection{Privacy guarantees offered by Aggregate Hints}\label{sec:privacy-guarantees-offered-by-aggregate-hints}
After the process step, all the transactions are sent to the searchers, who play the role of the adversary in the system, trying to learn about any individual transaction to frontrun it. Everything the matchmaker outputs is public and is the same for all searchers in the system. Not all users have to send a transaction to the matchmaker in every round, and all searchers can also input transactions to the matchmaker as a user.

As presented earlier, the matchmaker plays a crucial role in our design and
serves as a trusted curator for applying privacy mechanisms. While we
describe a system with a single point of trust in the system for the
sake of simplicity and to fit the current version of MEV-Share, a more
realistic deployment could see the matchmaker role being decentralized
and operated by various entities. In that case, the trust factor could
not be solely based on the operator's reputation, as is the case in the
present production deployment, but would have to be revisited to
integrate, for instance, a Trusted Execution Environment (TEE) that would
protect users' input privacy by offering cryptographic guarantees that
the matchmaker does not have access to the plaintext transactions.
Coupling Differential Privacy and cryptography to decentralize the role
of the Trusted Curator is a widely studied problem in the literature
\cite{chowdhury2020cryptepsilon,Cheu_2019,Bittau_2017}, and our approach could be combined with these
methods to remove its reliance on a Trusted Curator as well. This extension is out of the scope of this study, and we instead assume a trusted
matchmaker.

The guarantees our contributions offer to focus on offering users output
privacy on the hints the matchmaker exposes to searchers. With this
context in mind, we describe our system's formal privacy guarantees and start by presenting the threat model we are considering.

\subsubsection{Threat model}\label{sec:threat-model}

Our method aims to produce DP aggregate stats that can be shared as global hints to a MEV-Share-like system. Since our
work does not focus on a particular MEV-Share deployment or a
specific protocol, we do not delve into the specific DP settings that
would be needed to deploy the mechanism and instead assume a correct
implementation guaranteeing pre-trade privacy to users contributing
their transactions given a fixed privacy budget and correctly calculated
sensitivity for each of the DP mechanisms deployed. Instead, we focus on
what attackers can achieve thanks to the system being permissionless by
posing as regular users contributing transactions to the aggregate
hints.

Let's consider an example attack where malicious adversaries attempt to
break the privacy guarantees of a DP sum. In this scenario, we have a
matchmaking service that aggregates transaction values using
differential privacy to protect the privacy of individual transactions.
The system could be subject to
the following attack against the DP sum without adding random sampling. Legitimate users submit
transactions with valid and accurate values, ensuring the privacy of
their data through the differential privacy mechanism.
Malicious adversaries submit transactions with intentionally manipulated
values. For example, they could submit a large number of transactions
with extremely high values or very low values, aiming to bias the
overall sum. Per its differential privacy
mechanism, the matchmaking service adds controlled random noise to the sum of transaction values
to protect individual privacy. The amount of noise added is determined
by the privacy budget allocated for the aggregation. Since every transaction value directly influences the sum, the presence of
malicious attackers submitting transactions with extreme values can
impact the final result. If the attackers submit a substantial number of
transactions with inflated values, they can potentially overwhelm the
effect of the random noise and bias the sum towards higher values. In
the worst-case scenario, if the attackers' impact on the sum outweighs
the noise added by the differential privacy mechanism, it may be
possible for them to infer the true values of some transactions. By
comparing the released sum with the sum of the true values of their
submitted transactions, the attackers might deduce the individual
transaction values, thus breaking the differential privacy guarantees.

\subsubsection{Impact of the attack}\label{impact}

If a malicious attacker can submit enough transactions to make up the
majority of the transactions being considered for the aggregate stats,
it can potentially gain some information about individual transactions.
Depending on the operation computed, this would change the underlying
assumptions when calculating the operation's sensitivity and calibrating
the noise required to make the mechanism differentially private. This
type of attack of poisoning attack is particularly prevalent in the
Local Differential Privacy (LDP) model \cite{cheu2019manipulation} where it is challenging
to verify the integrity of users' noisy contributions.

In our case, the trusted curator's ability to mitigate the attacks would
be limited by the type of malicious adversaries we are considering.
Indeed, in a permissionless setting such as MEV-Share, nothing
prevents an adversary from trying and manipulating the aggregated results by
submitting a large number of transactions that match the predefined
criterion. Since the attacker controls a significant portion of the
transactions, it can impact the final aggregate statistics obtained by
the matchmaking service.

The consequences of such an attack could impact multiple actors in the
system. First, users would see a reduction in the privacy they expected when sharing their transactions with the matchmaker. The large number of transactions submitted by the malicious adversary might
outweigh the noise added through differential privacy mechanisms. As a
result, the aggregate statistics could become less private and reveal
patterns or characteristics of individual transactions.

Second, honest searchers would suffer from the adversary's influence on
the aggregated statistics since they would obtain distorted insights
about the state of the system, while the attacker would be able to
filter out the impact of its poisoning and, as a consequence, gain an unfair advantage over other searching bots who would make decisions
based on biased or manipulated information.

Finally, in a fully decentralized setting where multiple operators offer
the matchmaking service, a victim operator could be targeted by an adversary that would try to degrade the quality of the aggregate hints
it provides.

In all of the attack scenarios above, the cost for the attacker is the
same: while attackers would have to submit signed transactions, they can
astutely craft their attack such that most of these transactions end up
being canceled and do not incur extra costs to the attacker. More precisely, if the transactions the searcher submits have low gas fees and low MEV returns, the block builder would not prioritize it even though the transaction would still count towards the aggregation. Thus, the cost to the adversarial searcher to extract such information is not very large and, under ideal conditions, can be assumed to be 0.

\subsubsection{Attack mitigation}\label{attack-mitigation}

Typical Trusted Curator deployment would mitigate the threat model from
the previous section by rate-limiting users submitting an abnormal
number of transactions. However, in a permissionless system, the adversary can mount sybil attacks and spread its capital to appear as
many different users that would act in a coordinated fashion.

An option would consist in adjusting the mechanism's sensitivity. When
calculating the sensitivity of a mechanism, the matchmaking service can
take into account the presence of potentially malicious adversaries and
set a higher sensitivity value to add more noise to counter their
impact. Some DP statistics are more susceptible to being impacted by
malicious adversaries. DP statistics that directly depend on the full
range of the dataset, such as the mean and sum, can be more susceptible
to manipulation since every data point influences them in the
dataset, the presence of outliers or fake data points can significantly
affect the final result. This approach would degrade the overall utility
of the mechanism, without really discouraging attacks that aim to
degrade a service operator's performance.

We can, however, rely on the fact that the trusted curator acts as a relay
between the input data and the aggregate stats. Instead of including all
the transactions received during a period to publish the latest
batch of aggregate stats, the Trusted Curator will instead first sample
a subset of the available transactions and apply the DP mechanisms on
this subsample. By doing so, the matchmaker denies a malicious attacker
the certainty that his samples are included in the released stats.
The attacker cannot rely on prior knowledge to reverse
engineer the raw data used to compute the release DP stats. We will see
in the next section that this intuitive mechanism offers
greater privacy under approximate DP.

\subsection{Aggregate Queries}

Aggregate Queries leak two types of information to the searcher: Whether an aggregate condition has been fulfilled and the aggregation response. For the scope of the paper, we would assume that the aggregation conditions are a countOf query. When the countOf query satisfies a given condition, the aggregate statistic for similar transaction hints can be released. For the scope of this paper, we would be using Laplace noise, which provides us with pure differential (i.e., with $\delta = 0$). For now, the reason for choosing this is that the data is not a high-dimensional vector. Thus the benefits of using the $l2$-norm of sensitivity versus $l1$-norm of sensitivity are not substantial. 

 So why do we need differential privacy? One might argue that the matchmaker can keep the same aggregate (i.e., maintain a count and return whenever queried), implying the number of queries would not matter. However, the dataset is dynamic, and transactions are repeatedly added (when a new searcher inputs a transaction) and removed (when the transactions are bundled, sent to the matchmaker, and included on-chain).

\paragraph{Neighboring Inputs}

To maintain differential privacy, we need to ensure that for any neighboring datasets, the condition mentioned in Eq. \ref{eq:dpdef} needs to be satisfied. In that context, we first define what neighboring inputs ($X = (x_1, x_2, \cdots, x_n)$ and $X' = (x'_1, x'_2, \cdots x'_n)$ mean in this work.

$$X \sim X' \iff \exists x \in X ;\nexists x \in X'; \forall y \neq x (y\in X \iff y\in X')$$

Each $x \in X/X'$ are trades a user has sent to the matchmaker, and other trades already in the set have already satisfied the condition for differential private hints. The above equation can be said as "Datasets X and X' are neighboring, if and only if X contains only one additional entry compared to X'". Also, let $X-X'$ represent the trade that is in X but not in X'

% \paragraph{Random Sampling}

% Another aspect that needs to be addressed before we can finally define the queries is what all transactions are counted. For queries like sum, the sensitivity of the function is very high. However, it can be made even worse when an adversary specifically sends transactions (again no MEV extraction is possible in this transaction, and thus would not go on-chain, thereby no cost of attacking) to the matchmaker as a user. The adversary can set extremely high or low values in order to increase the sensitivity of the function. And then when we add laplacian noise to the aggregate, the noise would be too high to be meaningful for the searchers to reduce the search space. 

% To counter this, we introduce sampling before adding noise. This has an immediate effect on sensitivity since the outliers would only be selected with some probability. To achieve this, we can either sample randomly (uniformly), thereby only selecting outliers with a certain probability, or we can prune the outliers specifically so that they are never a part of the query. We adopt the former since as a searcher the most profitable bundles could be those that are in the outlier range, and knowing about them would help to get the best result. This parameter, however, we believe should be set by the searcher itself based on its searching mechanism.

\paragraph{Differential Privacy for countOf}

To respond to a countOf query (which can be in the condition or as an aggregate statistic response), we introduce the Laplacian noise into the COUNT query commonly used by database systems. Laplacian noise provides us with pure $\varepsilon$-DP guarantees and has the welcome property of being easier to compute. We already mentioned that the $l1$-norm of the sensitivity should not be a problem due to the low dimension of the database.

Next, we need to find the sensitivity of the function. This value is always 1 for counting queries: The final count can only change by 1 when a single user data is added or removed.

\paragraph{Differential Privacy for sumOf}

We introduce Laplacian noise to respond to a sumOf query (which can be only as an aggregate statistic response). We use Laplacian noise here, even though the sensitivity of the function is expected to be reasonably large. 

Next, we need to find the sensitivity of the function. For sum queries, there are two ways to maintain the sensitivity: either maintain the maximum traded amount after each trade comes in or estimate the maximum trade before a batch of trades. The former would require continuous count maintenance by the matchmaker, which could add to the latency. However, it would provide a better privacy guarantee. In the latter, we would be compromising on the sensitivity value, which, if incorrectly set, would lead to either more than expected privacy loss or more error than needed (based on whether we overestimated or underestimated the sensitivity). 

\paragraph{Impact of Random Sampling on Differential Privacy}

When implementing a differential privacy (DP) algorithm on a random subset of data, introducing uncertainty from the subset enhances privacy. This is because an individual's data may not be included in the analysis, thereby augmenting their privacy. Furthermore, the uncertainty regarding whether an individual's data was dropped or included contributes to privacy protection. The formalization of this intuition is provided by \textit{privacy amplification by subsampling} \cite{kasiviswanathan2011can,balle2018privacy}.

Privacy amplification by subsampling involves selecting a random subset of the original dataset and computing aggregate statistics solely on this reduced subset rather than the entire dataset. Consequently, the impact of individual data points on the released statistics becomes diluted, making it more challenging to deduce specific information about any individual's data. The underlying principle of privacy amplification lies in the decreased probability of any single data point being included in the subsampled dataset, thus reducing the risk of re-identification. As the size of the subsampled dataset increases, privacy guarantees improve. The subsampling lemma \cite{kasiviswanathan2011can} states the following:

Let $M$ be a $(\varepsilon, \delta)$-differential privacy mechanism, where $q = |U|/|N|$ represents the probability that an element $e$ will be included in a uniformly random subset $U$ drawn from a set $N$ with original elements. Then, $M_U$ is $(\varepsilon^{\prime}, \delta^{\prime})$-differential privacy, where $\delta^{\prime}=q \delta$ and $\varepsilon^{\prime}=\log (1+q(e^\varepsilon-1))=\mathcal{O}(q \varepsilon)$.

Therefore, mitigating the sybil attack against the aggregate DP hints has the welcome byproduct of increasing the overall privacy protection of the solution.

\section{Discussion}\label{sec:discussion}

\paragraph{Adoption of aggregate statistics in mev-share} Earlier in Section \ref{sec:search:bruteforce}, we described a strategy to cover the search space more widely by heuristically assigning the missing information in the trade. Here, we assumed heuristics exist based on external on-chain/real-world data. Aggregate statistics can serve as a boost to these heuristics and can be used to narrow the search space much further. This would lead to better bundles being submitted by the searchers without significantly improving the probability of unbundled sandwiches or frontrunning.
\paragraph{Experiment Results}
In the previous paragraph, we discussed that with a basic attack strategy like brute force search, the searcher can perform a better extraction with aggregate statistics. However, this does not apply to all attack strategies. Thus, we propose to publish the aggregate statistics as a dashboard for MEV-Share to allow the searchers to use and find more strategies that could involve using the aggregate data for more MEV.
\paragraph{On Privacy Budget selection}
The Trusted Curator selects the privacy budget for Differential Privacy based on the function and amount of data available. The privacy community considers that $\varepsilon < 1$ offers reasonable privacy protection. With sufficient input transactions, this should offer hints of sufficiently high quality for searchers to enhance their strategies with this extra information. The system's current design handles a single global privacy budget for all user transactions. Users contributing a large amount of data could benefit from setting a higher privacy budget to obtain greater rebates from searchers for the quality of the information they disclose. A natural extension to our system should thus consider these various categories of users and bucket them into different categories, each with a different privacy budget.

\section{Conclusion}

In this work, we introduce differentially-private aggregate hints for MEV-share. Our proposal could be implemented seamlessly in Flashbots' production infrastructure and be exposed as an opt-in feature for users. We expect searchers to benefit from this extra information and the new interactions with the matchmaker service. Our proposal comes at no extra privacy loss to the end-user and formally quantifies the amount of information leaked to searchers. 

We described the threat model of our protocol. We showed how random sampling could protect from sybil attacks and increase users' privacy while guaranteeing the quality of the aggregate hints provided to searchers in a permissionless environment.

In the next iteration of this work, we hope to be able to deploy our proposal to the main Flashbots' infrastructure and collect the relevant data required to quantify the utility gains our method offers searchers.

\newpage
\bibliographystyle{ACM-Reference-Format}
\bibliography{sample-base}
\vfill
\appendix

\section{Searcher strategies pseudo-code}

\begin{figure}[H]
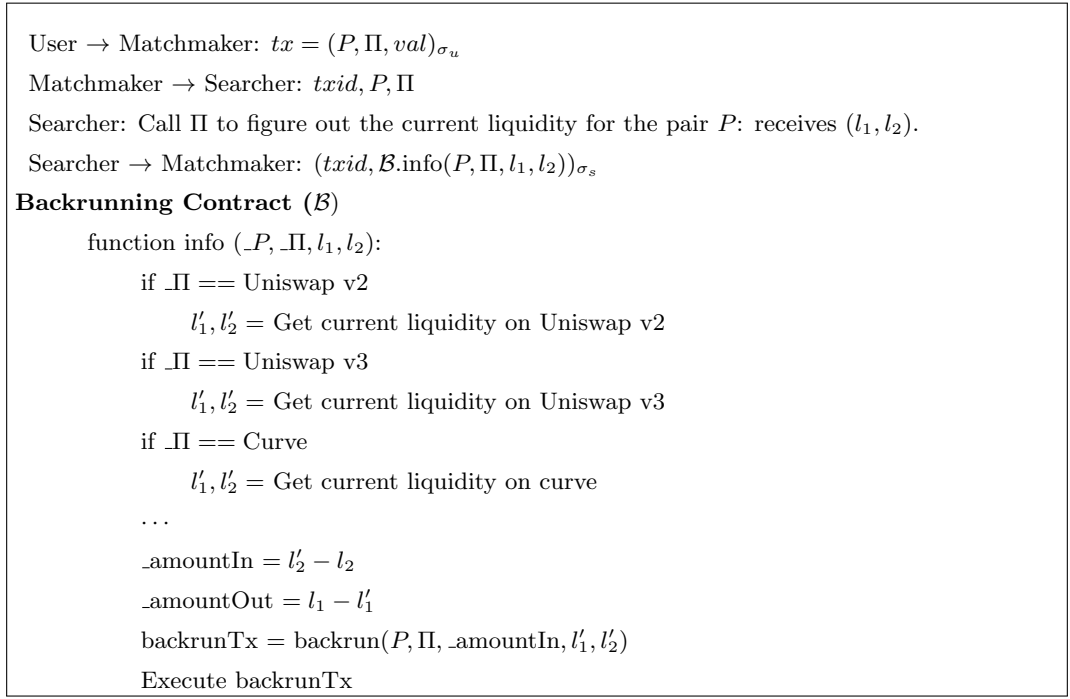

\begin{boxedminipage}[t]{\columnwidth}\small
\begin{description}
\item []
\item User $\rightarrow$ Matchmaker: 
 $ tx = (P,\protocol,val)_{\sigma_u}$
 \item Matchmaker $\rightarrow$ Searcher: $txid, P, \protocol$
 \item Searcher: Call $\protocol$ to figure out the current liquidity for the pair $P$: receives ($l_1,l_2)$.
 \item Searcher $\rightarrow$ Matchmaker: $(txid, \mathcal{B}.\text{info}(P,\protocol,l_1,l_2))_{\sigma_s}$

 \item[Backrunning Contract ($\mathcal{B})$] 
 \item[] 
\begin{itemize}
    \item [] function info ($\_P,\_\protocol,l_1,l_2$):
    \item [] if $\_\protocol == \text{Uniswap v2}$
    \begin{itemize}
        \item [] $l_1', l_2' = $ Get current liquidity on Uniswap v2
    \end{itemize}
    \item [] if $\_\protocol == \text{Uniswap v3}$
    \begin{itemize}
        \item [] $l_1', l_2' = $ Get current liquidity on Uniswap v3
    \end{itemize}
    \item [] if $\_\protocol == \text{Curve}$
    \begin{itemize}
        \item [] $l_1', l_2' = $ Get current liquidity on curve
    \end{itemize}
    \item [] $\cdots $
    \item []\_amountIn $= l_2'-l_2$
    \item []\_amountOut $= l_1-l_1'$
    \item []backrunTx = backrun$(P,\protocol,\_\text{amountIn}, l_1',l_2')$
    \item [] Execute backrunTx
\end{itemize} 
 
\end{description}
\end{boxedminipage}
\caption{Backrunning through contract example when searcher releases $P, \protocol$ assuming a function backrun which gets a backrunning bundle for a transaction when input token pair $P$, Protocol address $\protocol$, input amount $val$, and liquidity of the pair $(l_1,l_2)$.}
\label{fig:contractbackrunning}
\end{figure}

\begin{figure}[H]
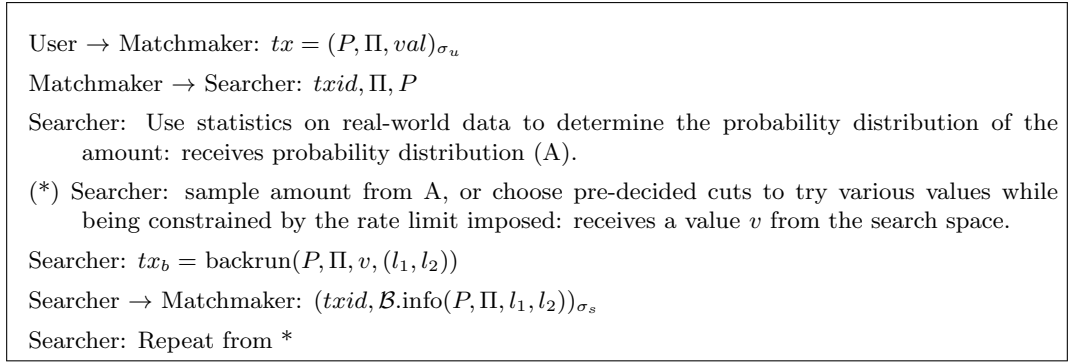

\begin{boxedminipage}[t]{\columnwidth}\small
\begin{description}
\item []
\item User $\rightarrow$ Matchmaker: 
 $ tx = (P,\protocol,val)_{\sigma_u}$
 \item Matchmaker $\rightarrow$ Searcher: $txid, \protocol, P$
 \item Searcher: Use statistics on real-world data to determine the probability distribution of the amount: receives probability distribution (A).
 \item (*) Searcher: sample amount from A, or choose pre-decided cuts to try various values while being constrained by the rate limit imposed: receives a value $v$ from the search space. 
 \item Searcher: $tx_b =$ backrun($P,\protocol, v,(l_1,l_2)$)
 \item Searcher $\rightarrow$ Matchmaker: $(txid, \mathcal{B}.\text{info}(P,\protocol,l_1,l_2))_{\sigma_s}$
 \item Searcher: Repeat from *
\end{description}
\end{boxedminipage}
\caption{Brute force based backrunning example when searcher releases $P, \protocol$ assuming a function backrun as in Fig \ref{fig:contractbackrunning}.}
\label{fig:bruteforcebackrunning}
\end{figure}

\section{Comparison of the two methods}
\label{comparison}

In practice, both methods have their advantages and disadvantages. While the first method can determine backrunning transactions accurately, it costs a lot more gas fees due to various on-chain calls to find the exact state of the contracts to calculate the backrun. The second method (brute force search) does not take too much gas cost, as most computation is done off-chain, but causes 1) sub-optimal backrun due to search inaccuracies and 2) Potential to spam the matchmaker and the block builder. Both these methods can be used in unison to get even better results. Let us say that the user opts not to reveal the protocol. Then since we know that Uniswap V2 is the most common trade, we can heuristically try backrunning the same. On the other hand, if the amount of trade is unknown, and the heuristics don't have a very high impact, then the value of the amount traded can be found in the backrun contract. However, in both these methods, the value extracted is compromised, but the privacy of the transaction is maintained in such a way that frontrunning is impossible.

% \subsection{Optimal Backrunning}

% Towards the other end of the spectrum would be the optimal backrunning strategy when all information about the transaction is disclosed. Here, since the transaction would be completely disclosed, the searchers can potentially try to sandwich or frontrun the transaction as well, but in order to compare the effectiveness of deferentially private hints, we will only consider the optimal backrunning profit from the same. 

% \section{Gaussian Noise}
% \label{gaussian}
% A noise that follows a Gaussian distribution with the following parameters.
% \begin{equation*}
%     f(x | \mu, \sigma^2) = \frac{1}{\sqrt{2\pi\sigma^2}} \exp\left(-\frac{(x-\mu)^2}{2\sigma^2}\right)
% \end{equation*}
% where $\Delta_2$ is the $l2-$sensitivity of the mechanism.
% The resultant mechanism is $(\varepsilon, \delta)-$Differently Private.

\end{document}